\documentclass[preprint,12pt]{elsarticle}

\usepackage{amssymb}
\usepackage[utf8]{inputenc}
\usepackage{lineno}
\usepackage[]{hyperref}
\usepackage{times}
\usepackage{graphicx}
\usepackage{latexsym}
\usepackage{alltt}
\usepackage{epsfig}
\usepackage{wrapfig}
\usepackage{amsmath}
\usepackage{float}
\usepackage{algorithm}
\usepackage[noend]{algpseudocode}
\usepackage{tabularx}
\usepackage{subcaption}

\journal{}

\begin{document}

\begin{frontmatter}

\title{Indexing Execution Patterns in Workflow Provenance Graphs through Generalized Trie Structures}
%
%

\author{Esteban García-Cuesta (esteban.garcia@universidadeuropea.es)} \fnref{myfootnote}
\address{Data Science Laboratory, School of Architecture, Engineering and Design, Universidad Europea de Madrid, C/Tajo S/N, Villaviciosa de Odón, 28670, Spain}
\author{José Manuel Gómez-Pérez (jmgomez@expertsystem.com)} 
\address{Expert System, Prof. Waksman 10, 28036 Madrid, Spain}

\fntext[myfootnote]{Corresponding Author}

\begin{abstract}
Over the last years, scientific workflows have become mature enough to be used in a production style. However, despite the increasing maturity, there is still a shortage of tools for searching, adapting, and reusing workflows that hinders a more generalized adoption by the scientific communities. Indeed, due to the limited availability of machine-readable scientific metadata and the heterogeneity of workflow specification formats and representations, new ways to leverage alternative sources of information that complement existing approaches are needed. In this paper we address such limitations by applying statistically enriched generalized trie structures to exploit workflow execution provenance information in order to assist the analysis, indexing and search of scientific workflows. Our method bridges the gap between the description of what a workflow is supposed to do according to its specification and related metadata and what it actually does as recorded in its provenance execution trace. In doing so, we also prove that the proposed method outperforms SPARQL 1.1 Property Paths for querying provenance graphs. 
\end{abstract}


\end{frontmatter}


\section{Introduction}
Scientific workflows have become well-known means to encode knowledge and experimental know-how~\cite{deelman2009workflows} in data-intensive science, playing an important role to make science repeatable and incremental, as well as to contribute to a better sharing, exchange and reuse of scientific methods and experimental outcomes. The abundance of workflow management systems like Taverna~\cite{doi:10.1093/nar/gkt328}, Kepler~\cite{Ludascher:2006:SWM:1148437.1148454}, Wings~\cite{Gil:2007:WPC:1620113.1620127} and others, and of popular public workflow repositories and e-Labs with thousands of users and hosted workflows, like myExperiment \cite{myexperiment}, CrowdLabs \cite{crowdlabs11} or Galaxy~\cite{gal11}, witnesses such development and shows the extent to which community efforts have focused in this direction.

However, despite the increasing degree of maturity displayed by the different scientific workflow management systems and platforms, general adoption still seems to be under way. Domain scientists tend to perceive scientific workflows  as powerful computational tools, but also as complex and rigid artifacts of difficult practical application. Similarly, potential adopters in the scientific domain with advanced programming skills may prefer to process data in their usual programming languages, like R, Python or MatLab, due to the existence of large user communities and the availability of well organized user support, ad-hoc libraries and a preexisting body of work. As a consequence, persuading data-intensive scientists and related user communities to adopt scientific workflows is not an easy task. Actually, there seems to be a consensus that a wider adoption of scientific workflows will only be achieved if the focus of research and development is shifted from methods and infrastructure for developing and executing workflows to better means to search, adapt, and reuse them~\cite{cohen11}. 
 
This requires more expressive ways to describe scientific workflows, e.g. through richer metadata and workflow description languages. However, since scientific workflows are designed specifically to compose, orchestrate and execute a series of computational or data manipulation steps in a scientific application, we find that describing a workflow is a multifaceted task that involves its specification, i.e. what the workflow is supposed to do, but also what it actually does as evidence is recorded that characterizes the workflow from an execution point of view. In this paper, we focus on the latter and present the following main contributions:
\begin{itemize}
\item A set of \textbf{workflow management functionalities, including indexing, search, and similarity detection based on the analysis of the provenance graphs} produced during scientific workflow executions. Our approach is based on the application of data mining techniques to enable the discovery of sequential or linked patterns in the provenance graphs, allowing the description of widely-used relations among objects in the execution trace of a workflow that may or may not be defined formally, e.g. though ontologies~\cite{Getoor03,zhang12}.
\item To this purpose, we propose the use of \textbf{statistically enriched generalized trie structures}~\cite{knuth97}\cite{fredkin60} focused on indexing arbitrary data types, supporting efficient sequence mining and analysis, and identifying deterministic paths with no need for dynamic reorganization.The generalized trie structure allows to find the commonly linked processes within workflow executions and facilitates access and retrieval by generating a dedicated index for both Direct Graphs (DG) and Direct Acyclic Graphs (DAG).
\item As a corollary to our approach, our method is also useful to \textbf{alleviate the efficiency issue of SPARQL 1.1 property paths in RDF graphs} by generating, in batch mode, an indexing structure and allowing incremental updates (both structure and statistics) as new data is added to the repository.
\end{itemize}   


The remainder of this paper is organized as follows: Section~\ref{sec:rel} provides an overview of the related work and state of art of the proposed indexing, mining, and querying functionalities. The details and description of the statistically enriched generalized trie is presented in section~\ref{sec:indexing}, including the algorithms and implementation details. In section~\ref{sec:experiments}, we describe the datasets and the experiments designed for the evaluation of our approach. Section~\ref{sec:results} presents the results of the experiments evaluating both the extended workflow management functionalities and the time performance analysis, including a comparative study against the SPARQL 1.1 Property Paths W3C specification. Finally, we conclude the paper and introduce future work in section~\ref{sec:conclusions}.

\section{Related Work}
\label{sec:rel}
Due to their complex nature, a major difficulty in scientific workflow management lies in properly describing scientific workflows in meaningful ways that are also readable both by humans and machines. As a matter of fact, there is good progress in this regard, e.g. through the development of interoperable workflow modeling languages like CWL \cite{de702af8-ea61-47af-aabb-286b2274e1be}. Nevertheless, the lack of descriptive metadata should be addressed not only through more expressive, interoperable workflow description languages but also by actually enriching the description of the artifacts involved in the actual computation. 

However, adding such metadata usually becomes labor-intensive, and ultimately scarce. Research objects \cite{Bechhofer2013599} provide containers of scientific knowledge, i.e. semantically rich aggregations of resources that bring together the data, methods and people involved in a scientific investigation as a single information unit. In \cite{8109145}, the application of natural language processing and semantic annotation technologies allowed the  automatic generation of metadata from the payload of research objects contained in ROHub\footnote{http://www.rohub.org}, the reference platform for research object management \cite{palma2014rohub}, including resources such as scientific publications, technical reports and presentations and textual workflow descriptions contained in their specification. As a result, richer, self-descriptive, expressive and machine-processable research objects were produced while reducing human annotation effort. Workflow-centric research objects also encapsulate all the necessary metadata to preserve scientific work against potential decay \cite{Zhao2012WhyWB}, including workflow execution provenance. 

Thoroughly characterizing a scientific workflow requires a twofold strategy. We certainly need to take into account the information related to its specification and related resources, but, just like other types of software, this provides a partial understanding of the code. Only through the analysis of its execution will we have access to the necessary insight to understand how a workflow works, debug it if required, and store it in a repository with other, functionally similar, software components. In this regard, we distinguish between a workflow specification, in some kind of modeling language~\cite{garijo2014common}, and the provenance trace resulting from the execution of the specified workflow. Such provenance~\cite{moreau10} traces contain information about each of the steps in the data transformation sequence and keep track of all the computations performed during the execution of the workflow. 

Related process mining work~\cite{1316839} developed techniques and implemented toolkits like ProM \footnote{http://www.processmining.org} for discovering and visualizing workflow models from execution logs. However, approaches for managing provenance differ greatly and the concrete relation between a workflow and its traces varies significantly due to the differences in the semantics of workflow specifications~\cite{mor08}~\cite{cohen09}. In \cite{Alper:2014:LEW:2977935.2977945}, the authors enrich workflow provenance with additional annotations so that provenance can be utilized to label various data artifacts, alleviating the lack of domain-specific metadata in provenance traces in support of tasks like reporting  scientific workflow computations. In \cite{garijo2014common}, the authors presented a study of common motifs, understood as common execution patterns related to workflow execution provenance, and introduce the convenience of identifying such motifs to cluster related workflows from an execution point of view. This is known to be a hard problem and, in its more general form, it is equivalent to searching topologically identical subgraphs, which is considered NP-complete. Other approaches~\cite{6655668} focus on enabling an efficient and scalable storage and querying of large collections of provenance graphs serialized as RDF graphs through distributed big data stores like Apache HBase~\footnote{https://hbase.apache.org}. These combine ad-hoc indexes over the store with evaluation algorithms that rely on such indexes to compute expensive join operations. Albeit scalable and highly performing, these approaches are also limited since they are unable to identify common motifs or execution patterns at higher levels of abstraction, like the ones above-mentioned, which enable the clustering of similar workflows in terms of their execution behavior. Overall, it can be concluded that most of the existing management systems and content analysis processes treat workflows as atomic entities and make use of indexing processes based on keywords~\cite{cohen11}.  

Alternatively, we consider a workflow as a set of linked processes that can be analyzed in order to obtain a deeper knowledge of its execution, behavior and objectives. There are some ongoing research initiatives that follow this approach, in an attempt to extract patterns by automatically learning from the interlinked structures. In~\cite{zhang12} the authors refer to this process as link pattern discovery being its main goal the characterization of interlinked relations which may or may not be defined explicitly. Also, at~\cite{Getoor03}\cite{Getoor05} the authors referred to the term link mining in order to indicate how data mining techniques can explicitly consider the links between objects when building predictive or descriptive models of linked data. Notice that the analysis of interlinked structures are also applicable to other relevant open datasets such as DBpedia\footnote{http://dbpedia.org/}.

In this experimental scientific context, index structures are core components and the current literature classifies them into three main categories based on the type of structures they use: i) tree-based structures, ii) hash-based techniques, and iii) trie-based structures. The first two are commonly used for general purpose indexing, while the third was originally designed and mainly used for string operations~\cite{boehm11}. However, the versatility of trie structures enables different purposes. For instance, trie structures have been recently used successfully in applications including efficient access to XML improving query performance~\cite{brenes08}, the representation of documents in a clustering application~\cite{morariu11}, or in a user modeling, where they enable the generation of a bigger new set of temporal user's features~\cite{DBLP:journals/eswa/Garcia-CuestaI12}. 

In order to find workflow patterns, the index structure must capture syntactic and semantic aspects represented as a graph. At~\cite{Grig08} the authors describe a graph matching approach to match behavioral descriptions of services, but the method does not scale to large repositories and does not provide approximate matching either. Despite there exists approaches in the graph search community for solving approximate matching~\cite{he08}, as far as the authors know, they have not been applied successfully in the domain of scientific workflows yet.

Thereof, the use of an index as a way of simplifying the graph matching problem into an approximate matching one provides a twofold solution for gaining efficiency in the search and also potentiality in the pattern discovery process. Such simplification is sometimes also preferred since more information, such as what might be missing or spurious in a query of a database graph, can be captured too. The proposed statistically enriched generalized trie supports the characterization of scientific workflows based on their execution behavior in the presence of very large (provenance) graphs. The generalization is based on prefix trees (PT)~\cite{fredkin60} --also called digital search trees~\cite{knuth97}-- to use parts of the key to determine the path within the trie. The prefix trees are mainly used for string indexing, where each node holds an array of references according to the used character alphabet~\cite{heinz02}. The generalization of this structure allows to maintain a tree of ordered data types where each node has at least two children and each edge encodes the specific data type. On the downside, this structure requires exponential allocation storage, and was consequently neglected for in-memory indexing of arbitrary data types~\cite{tobin86}. However, this main drawback has been mitigated due to the recent technological advances allowing larger memory size at cheaper prices and there also exist storage efficient alternatives implementations~\cite{boehm11} , making it suitable for new applications as proposed here.

\begin{figure}[ht]
\centering
\epsfig{file=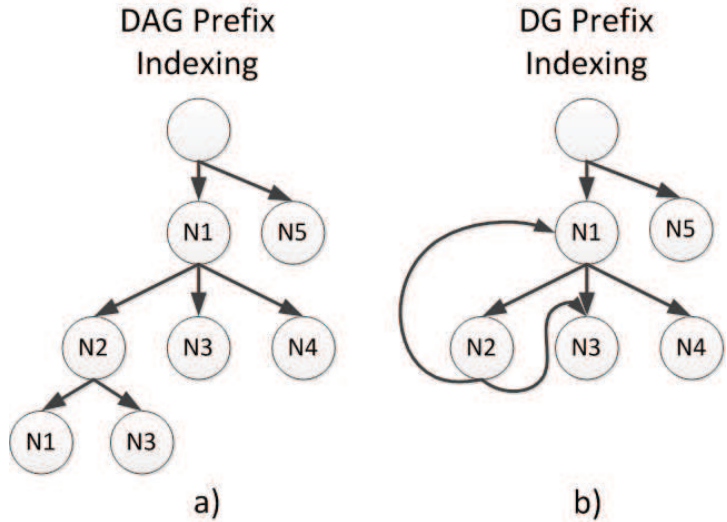, scale=0.41}
\caption{ The indexed sequences are: \{\{N1,N2,N1\},\{N1,N2,N3\},\{N1,N3\},\{N1,N4\},\{N5\}\}. a) Shows the resulting prefix indexing on DAG. b) Shows the resulting prefix indexing on DG.}
\label{fig:prefixTree}
\end{figure}

Figure~\ref{fig:prefixTree} shows the DAG and DG Prefix Tries generated given a set of sequences. These indexing algorithms are related with the two scenarios that we are testing: i) mining provenance of scientific workflows for finding similarities and assist users during the designing phase (DAG), and ii) mining general RDF graphs such as DBpedia (DG). These two indexing algorithms are two different versions of the prefix tree structure including the canonical label definition during the preprocessing step which is previous to the indexing itself. The first is a generalized prefix tree that dynamically stores the keys of a direct acyclic graph (DAG), whereas the second supports not only acyclic graphs but any directed graph (DG) (allowing cycles) by considering self-references whenever two nodes have the same identifier.

On the other hand, related graph management needs in the Semantic Web community inspired the integration of property path capabilities for SPARQL 1.1\footnote{http://www.w3.org/TR/sparql11-query/}, under the auspices of the W3C. Despite this effort, it has been proved empirically \cite{marcelo2012} that the use of property paths, e.g. to look up FOAF chains, was not possible in a reasonable amount of time, even in small RDF graphs and with very simple property path expressions. According to such studies, this is not due to particular implementation issues but to the SPARQL 1.1 specification itself, the reason being the need for counting solutions imposed by the specification. Therefore, even for simplified path problems that do not contain cycles, algorithmic complexity is still \#P. Our approach can also be used for converting a RDF graph into a trie structure begin a possible solution to this problem as shown in Section~\ref{sec:results}. Notice that most of the existing graph matching methods are not applicable for very large graphs and by extension either is possible to query them (this occurs with SPARQL 1.1. property paths~\cite{marcelo2012}).

\section{Indexing, Querying, and Mining Provenance}\label{sec:indexing}

Being able to extract relevant information and knowledge from the provenance of workflow executions, requires to analyze and transform the data into a suitable and machine-interpretable form. This provenance mining process is composed by three main steps: i) preprocessing and indexing, ii) querying and mining workflow executions, and iii) and interpretation and applicability. Our approach to this process is based on the definition of canonical label introduced by gSpan algorithm~\cite{xifeng02}, which establishes that any graph can be unambiguously mapped to a tree by using a combination of: lexicographical ordering (i.e. 'A' $<$ 'B' $<$ 'C' $\ldots$ $<$ 'Z') and minimum Deep First Search (DFS) code selection, and the inclusion of statistical information in the nodes of the trie. The approach is described in the following.

\subsection{Preprocessing and indexing algorithm description}
The provenance of workflow run $provwr$ is defined as a DAG that contains a set of resources $R$ described by a unique identifier (Uniform Resource Identifier URI) and ordered by its links: $provwr = r \cup e$, being $R = \{r_{i}, r_{o}, r_{p}\}$ the set of available resources where $r_{i}$ are input resources, $r_{o}$ are output resources, $r_{p}$ the process resources, and $e_{i,j} = (r_{i}, r_{j}) \, \forall i,j : i,j \in {R}$ the set of edges $E$ that links two different resources of $R$. In our scenario the provenance information is described by the PROV-O ontology\footnote{http://www.w3.org/TR/prov-o/} and the objects are implicitly defined by RDF graphs. Thereof, the only mode for accessing to the different sequences of execution is determined by reasoning according to RDF semantics~\cite{hayes04}. Hence, it is needed to transform the graphs $G$ into sequential path structures $T$ which can be inserted into the described prefix trie index by performing an intermediate step. This transformation $f[u]: G \to T$ from unordered provenance of workflows to a topological ordered sequence has been done by applying a Depth-First Search (DFS) + resource lexicographical ordering (LEX), which establishes a unique and deterministic ordered method whenever there are two or more possibilities for choosing the next process during the sequencing transformation $f[u]$. Then, two partially ordered sets $A$ and $B$ are defined as$(a,b) \leq (a',b')$ if and only if $a < a'$ (or $a=a'$ and $b \leq b'$). Note that in order to allow searching for similarities which are covered only partially by a workflow we have also allowed as an option the indexing of sub-sequences produced by this intermediate step by creating n-processes sequences (where $n$ defines the length of the sub-sequence) such as given $r_{i} = \{r_{i1},r_{i2},r_{i3},r_{i4}\}$ and a value of $n=3$ the following 3-resources are indexed: 3-r$_{i}= \{ \{r_{i1},r_{i2},r_{i3}\}, \{r_{i2},r_{i3},r_{i4}\} \}$.

\begin{algorithm}
\caption{Prefix-Tree provenance index algorithm}\label{alg:euclid}
\label{prefixTreeAlgorithm}
\begin{algorithmic}[1]
\Require a DAG provenance path $provwr = \{r_1,\ldots, r_s\}.$       
\Require the n-value of the sub-sequence size
\Procedure{InsertTrie}{$prowr, PT$}  \Comment{PT: Root Node of the Prefix-Trie structure}
\For{\textbf{each} $\{r_1,\ldots,r_k\} \in provwr_k \in provwr$} 
\State currentPT = PT
\State depth = 0; i=1 
\While{$((r_i == r_i.next())$ != empty)} \{
\State {URI = getURI($r_i$)} 
\State depth++
\If{(currentPT.getChild(URI, depth) == TRUE)} \{
\State currentPT =  currentPT.getChild(URI))
\State currentPT.freq++
\State currentPT.updateStats() 
\State \}
\Else \{
\State create new node N 
\State N.freq++; N.prob=1
\State currentPT.addChild(N)
\State currentPT.updateStats()
\State currentPT=N 
\EndIf \}
\EndWhile \}
\EndFor
\State \textbf{return} $PT$ \Comment{a Prefix-Trie PT that records all the prefixes of P and statistics}
\EndProcedure
\end{algorithmic}
\end{algorithm}

\begin{figure}[ht]
\centering
\epsfig{file=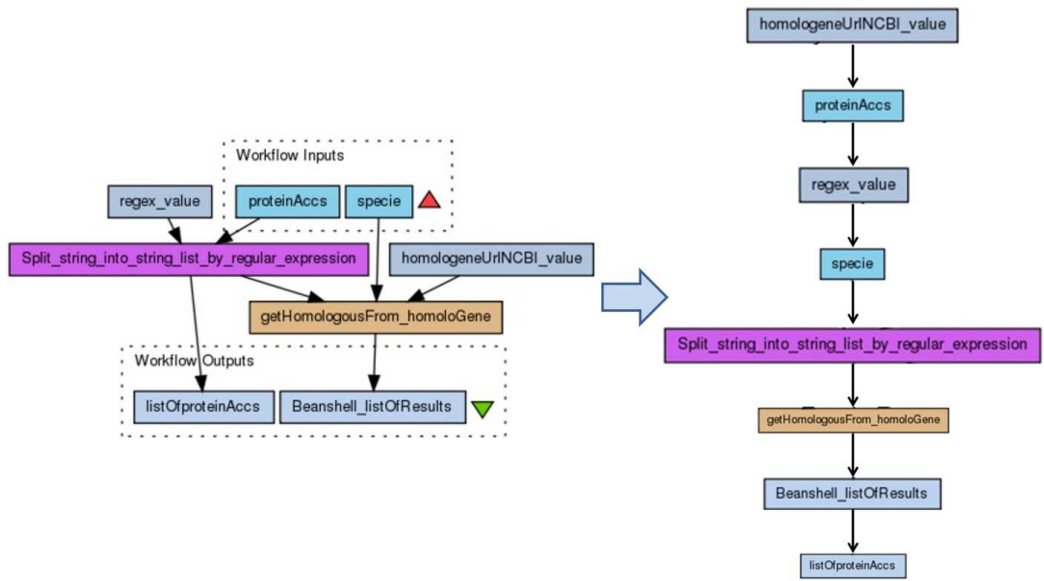, scale=0.28}
\caption{DFS+lexicographical ordered "Get homologous from NCBI homoloGene" workflow obtained from MyExperiment.}
\label{fig:sorted}
\end{figure}

This approach avoids to have several different ordered sequences for the same DGA. For instance, two input resources $r_{i1}$ and $r_{i2}$ of a process resource $ r_{p1}$ that returns $r_{o1}$ could be ordered as $r_{i1} \rightarrow r_{i2} \to r_{p1} \to r_{o1}$ or $r_{i2} \to r_{i1} \to r_{p1} \to r_{o1}$ and only the partial order between the vertex of the DAG which are linked can be assured; but, by using the same criteria during the whole sequencing transformation process, we are able to compare the different workflows and find similarities without losing information despite the original representation is not unique. This approach provides a unique map from a DAG to a sequence, being $f[u]$ a non-bijective mapping function using only the identifiers associated to the nodes. Otherwise, adding edge information would be needed. Figure~\ref{fig:sorted} also shows an example of application of the indexing algorithm to a real workflow. In the left the original workflow is shown which has a graph structure with multiple possible sequential representations and in the right the indexed workflow is shown after applying the DFS+LEX method. 

The algorithm's pseudo-code for inserting a given provenance of workflow run $provwr$ is shown at algorithm~\ref{alg:euclid}. The InsertTrie function receives a workflow provenance $provwr$ resulting from the application of the DFS+LEX above explained method and creates $k = s-n+1$ sub-sequences $provwr_k$ which are inserted into the tree. The identifier for each node is the URI of the process and it is used for indexing the different parts of the workflow execution . For each node,  the statistics related to frequency and probability of occurrence given the previous processes $P(r_{ik} | r_{i1} \ldots r_{i(k-1)})$ are updated and calculated at depth level in order to obtain frequent patterns and provide the most probable next step as a recommendation.

All the functionalities presented in this work focus on the indexation process rather than applications that need the reconstruction of the original information (e.g. information compression). Therefore, the inverse mapping $f[u]^{-1}$ which relates one $t \in T$ to many $g \in G$ does not imply any drawback (on the contrary, it naturally groups the data, which can be beneficial to improve access time performance~\cite{morariu11}). 

\subsection{Querying and mining workflow executions}
Applications in the scope of this work include e.g. the following scenarios: i) the discovery of possible workflow solutions to a given problem by defining the inputs and expected outputs, ii) the assistive design of scientific experiments for helping end-users to have a n-step ahead prediction, and iii) the indexing of most frequent in-use patterns. Other applications such as comparing runs to find out Points-of-Deviation for understanding why one workflow execution resulted in a different output than the execution of an identical workflow on the same data have also been pointed out at~\cite{cohen11}, though we do not cover them in this paper.

All such applications can be expressed as alternatives of searching in the indexing prefix tree structure and defined as the problem of finding all the possible alternative paths given a set of inputs and a set of outputs. The similarity workflow query is defined as the set of alternative paths from an initial point $A$ to a final point $B$ being $A=\{a^{(1)}, a^{(2)},\ldots,a^{(k)}\}$ and $B=\{b^{(k+j)}\}$ having $j-k$ wildcards\footnote{A wildcard is defined with the symbol '*' meaning any possible value of the alphabet $\Sigma$.}. Similarly, the assistive design functionality can be expressed as searching for all the alternative paths between A and any possible final resource B, given that $A=\{a^{(1)}, a^{(2)},\ldots,a^{(k)}\}$ and $B=\{b^{(k+j)}\} : \forall b \in R$. Last, the most frequent pattern query is accomplished by collecting the statistics of frequency and probabilities of occurrence for all nodes of the tree and also per level allowing to perform queries such as, what is the most probable node in a $j-k$ ahead step given $a^{(1)}, a^{(2)},\ldots,a^{(k)}$? This process of collecting and updating the tree is done in real time automatically every time a new sequence is added (see Algorithm~\ref{alg:euclid}). 

\begin{figure}[ht]
\centering
\epsfig{file=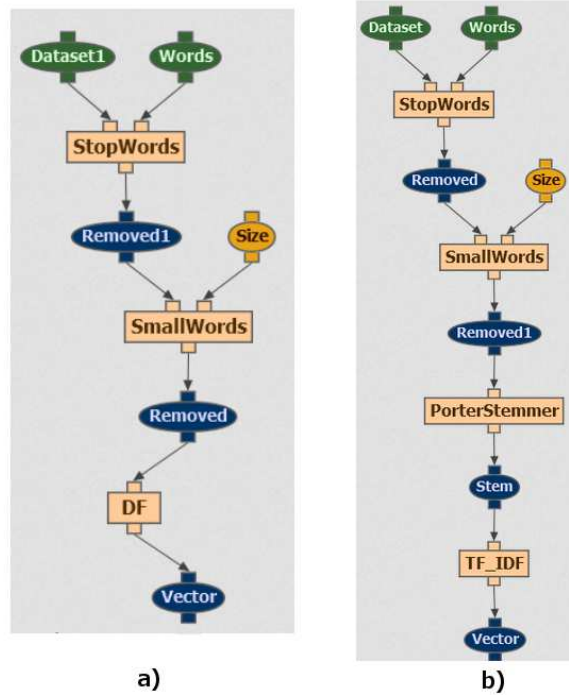, scale=0.40}
\caption{Workflows taken from Wings for illustating the similarity search alternatives problem: a)SimilarWords workflow; b)DocumentClustering workflow}
\label{fig:motivation}
\end{figure}

\begin{quote} \textbf{Illustration}: Let us consider that our user is working on a text mining application for clustering documents. One of the common processes is to remove stop words and apply some weighting step in order to find out the most relevant words. There are different steps that  may be applied to this purpose, ranging from the simple removal of a list of stop words to more complex approaches based on ontologies. Our user implements a preprocessing step as the one shown at Figure~\ref{fig:motivation} a)\footnote{http://www.opmw.org/export/page/resource/WorkflowExecutionAccount/ACCOUNT1348621567824} but she would like to get different possible alternatives for solving the same problem based on the kind of input/outputs. One possible suggestion would be the workflow shown at Figure~\ref{fig:motivation} b)\footnote{http://www.opmw.org/export/page/resource/WorkflowExecutionAccount/ACCOUNT1348621400515} where the "PorterStemmer" process is an extra step for removing commoner morphological and inflexional endings from words which may be useful to be incorporated at workflow a).
\end{quote}

The search for all the alternative paths querying process makes a traversal search over the tree using the node identifiers (URIs) as keys for finding temporal patterns that matches with the specified inputs and outputs. Such lookup process is at most linear over the size of the alphabet $\Sigma$ (in our case composed by $R= \{\textbf{r}_{i}, \textbf{r}_{o}, \textbf{r}_{p}\})$ and the depth of the DAG/DG trie structures $d$ being its complexity $O(\Sigma + d)$ . It is important to point out that for the construction of the query $q$ there is a need of transforming the DAG to be queried into a sequential ordered path using $f[u]$. As result, we obtain a new mapped query $q' = f[q]$ that can be searched within the prefix tree index. 

The two queries that we have implemented to show the capabilities of our approach and for validation of the scenarios i and ii introduced above in this section are: 
\begin{itemize}
\item \textbf{Q1} relates to the discovery of similar scientific workflows based on their provenance. $Q1M$ is defined as an approximate matching problem by introducing $M$ wildcards, where each wildcard represents a possible single walk from a starting resource $r^{(1)}$ to a final resource $r^{(2)}$.
\item \textbf{Q2}, a specific case of Q1 focused on the assistive design scenario, uses the statistics obtained during the indexing process to calculate the maximum likelihood given a current resource with the next $j$-resource, which can also be defined by a set of $r^{(1)}, r^{(2)}, \ldots,r^{(k)}$ followed by $j-k$ wildcards.
\end{itemize}

\section{Experimental Design}\label{sec:experiments}
We have performed two different experiments in order to validate the proposed method on graph searching and matching scenarios. The first validates the extended workflow management functionalities by checking that the number of obtained solutions are the theoretically expected for all the datasets and they are also retrieved on real time ($<250ms$). The second assess performance by measuring the performance benchmarking our approach against W3C SPARQL 1.1 Property Paths.  

The first experiment (E1) defines and executes Q1 queries (notice that Q1 query is an specification of the Q2 type having one less wildcard due to the last resource is set and hence both are equivalent for our validation purposes). Q1 query searches for workflows that have a specific input ($R_{i}$) and output ($R_{o}$), retrieving all the alternatives (set of random walk processes) that allow reaching the output given such inputs. In the second experiment (E2) we compare the time performance obtained by our approach (PT + DFS +LEX)  versus SPARQL 1.1. Property Paths. This evaluation measures the time needed for retrieving similar paths including wildcards as free nodes as specified by the Q1M query. An example of this types of queries expressed in SPARQL 1.1. syntax is: 'SELECT * WHERE \{ :r0 (:p)* :r1\}', replicating the property path '(:p)*' as many times as the number of needed wildcards.

For the validation of this two scenarios we have used three different datasets: i) the ProvBench corpus\footnote{https://github.com/wf4ever/provenance-corpus}~\cite{kha13} based on Taverna~\cite{miss10} and Wings~\cite{gil11}, and two clique datasets (complete undirected graphs) ii) of size 4 (4-clique), and iii) of size 8 (8-clique). The ProvBench corpus contains 120 provenance traces of workflow results related to 12 different application domains (text analytics, genomics, machine learning, social network analysis, proteomics, domain indepents, chemioinformatics, astronomy, heliophysics, phylogenetic, biodiversity, and weather forecast) and is licensed as Creative Commons. The clique graphs have been created as complete graphs in the same way as in~\cite{marcelo2012}.

All the experimentation have been performed over the three above-mentioned datasets and using an implementation on Java for our proposal and the usage of Jena-ARQv2.9.4 for SPARQL 1.1. property paths queries. Both methods have been built on a JavaSE System Library 1.7 and run in an Intel Core i5-3360M CPU, 2.80 GHz, 8GB RAM using Windows 7 64bits OS.

\section{Results and Performance}\label{sec:results}
Table~\ref{table:success} shows the E1 results obtained by executing the query Q1M for different number of wildcards and the different datasets using the proposed method (PT + DFS +LEX). The purpose of this query is to verify that the number of obtained solutions are the expected for all cases. Note that given a complete graph n-clique, the number of alternative paths (P) from a node A to a different vertex B in m steps is:  $P(A,B) = \frac{(n-1)^m-(-1)^m)}{n}$ (i.e. for $m=6$ and 4-clique graph $P=547$), and $\forall i,j : i \ne j \, \quad P(A_{i},B_{j})=\binom{n}{2} \frac{(n-1)^m-(-1)^m)}{n}$ (i.e. for $m=6$ and 4-clique graph $P=98460$). E.g. E.g. the second row and first column cell of Table~\ref{table:success} shows the number of alternative paths between any two different nodes for a 4-clique graph which are at distance 1 from each other and the query execution time. For that case the possible paths between two nodes $r^{(1)}$ and $r^{(4)}$ are: $\{r^{(1)},r^{(2)},r^{(4)}\},\{r^{(1)},r^{(3)},r^{(4)}\}$ as indicated, and the time needed for the execution of the query took less than 1ms.

We also tested the ProvBench repository by finding workflows that have as input the resource "\#dataset"\footnote{http://www.isi.edu/dc/TextAnalytics/ontology.owl\#Dataset} and "\#vector"\footnote{http://www.isi.edu/dc/TextAnalytics/ontology.owl\#Vector} as output and successfully checked that the number of provided solutions were the expected. Furthermore, the different columns of the table~\ref{table:success} show the results obtained for different number of wildcards (M) corresponding also to different the number of possible random walks between queried input $R^{(1)}$ and output resource $R^{(k)}$. 

For E2 validation we have tested the proposed indexing method (PT+DFS+LEX) versus the SPARQL 1.1. Property Paths recommendation, solving some of the problems related with computational efficiency that were showed up by~\cite{marcelo2012}. 

\begin{table*}
\centering
\begin{tabular}{|c|c|c|c|}
\hline
\multicolumn{4}{|c|}{Functionality Evaluation}\\\hline
\hline
                  & NºPaths/Time  & NºPaths/Time & NºPaths/Time \\
									& Q11 & Q12 & Q16 
 \\\hline  
ProvBench         & 3/$<$1ms &  22/$<$1ms &  2953/9ms    \\\hline
Complete 4-clique & 2/$<$1ms &  7/$<$1ms  & 547/9ms      \\\hline
Complete 8-clique & 6/$<$1ms &  43/1ms    & 102943/93ms  \\\hline
\end{tabular}\\
\caption{\small E1 experiment results. PT + DFS +LEX functionality evaluation for discovery of similar provenance query $Q1M$ for $M=\{1,2,6\}$.}
\label{table:success}
\end{table*}

\begin{figure}
    \centering
    \begin{subfigure}[b]{0.45\textwidth}
        \includegraphics[width=\textwidth]{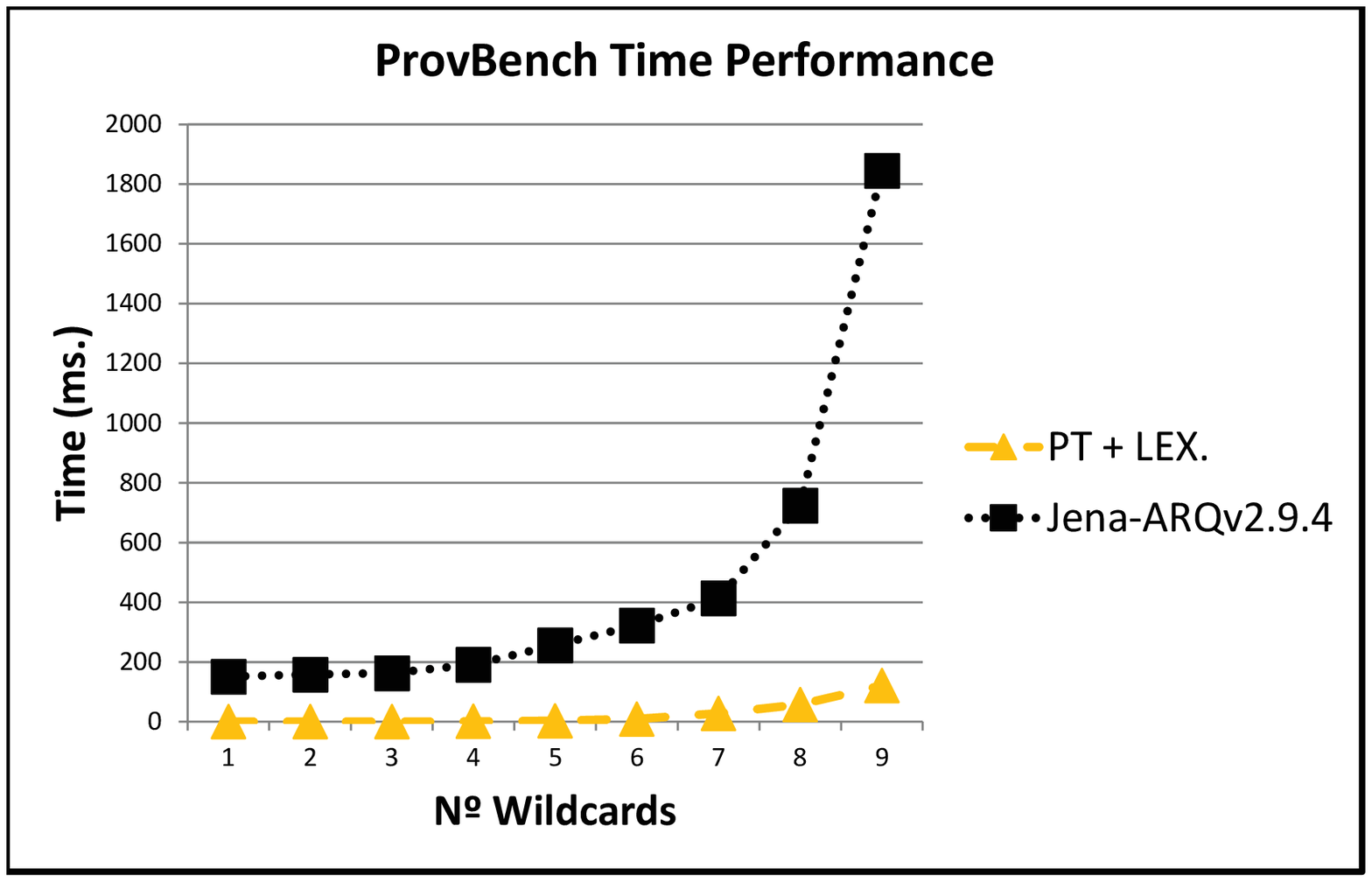}
        \caption{Time performance lookup operation of Provbench dataset.}
        \label{fig:timeProvBench}
    \end{subfigure}
    ~ 
    \begin{subfigure}[b]{0.45\textwidth}
        \includegraphics[width=\textwidth]{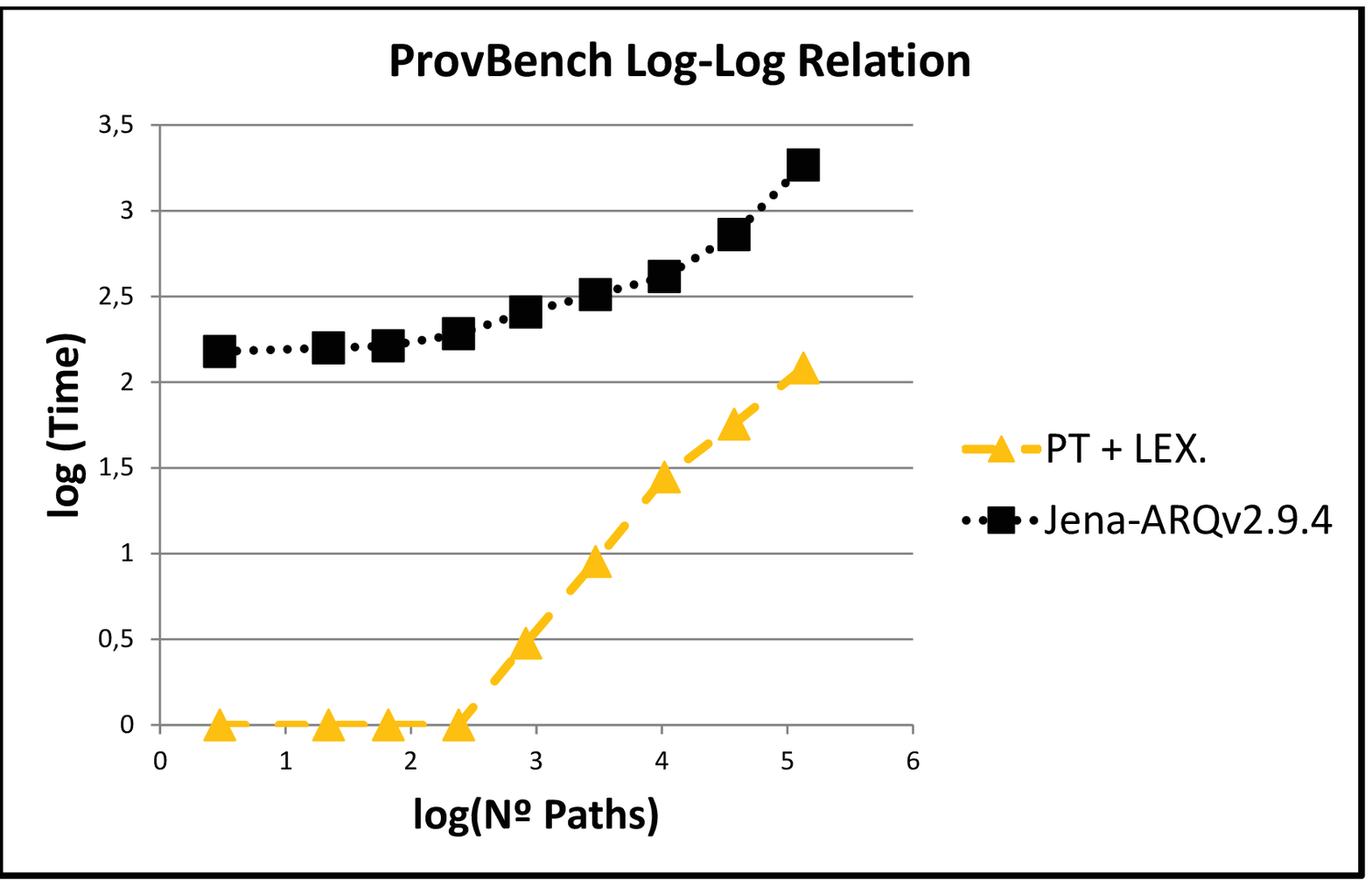}
        \caption{Time vs. number of paths log-log relation of Provbench dataset.}
        \label{fig:loglogProvBench}
    \end{subfigure}
    ~
    \begin{subfigure}[b]{0.45\textwidth}
        \includegraphics[width=\textwidth]{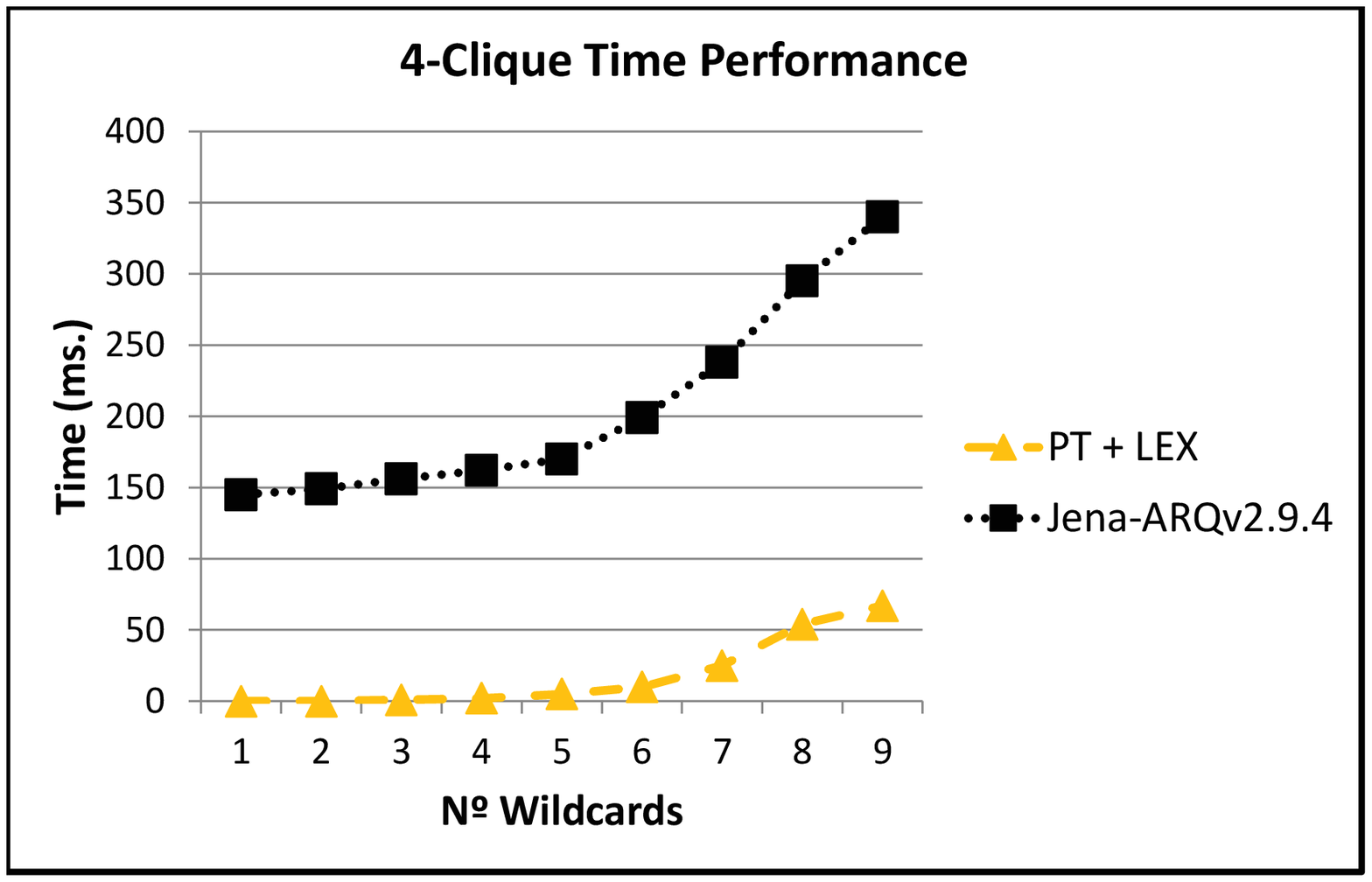}
        \caption{Time performance lookup operation of 4-clique dataset.}
        \label{fig:time4-clique}
    \end{subfigure}
    ~
    \begin{subfigure}[b]{0.45\textwidth}
        \includegraphics[width=\textwidth]{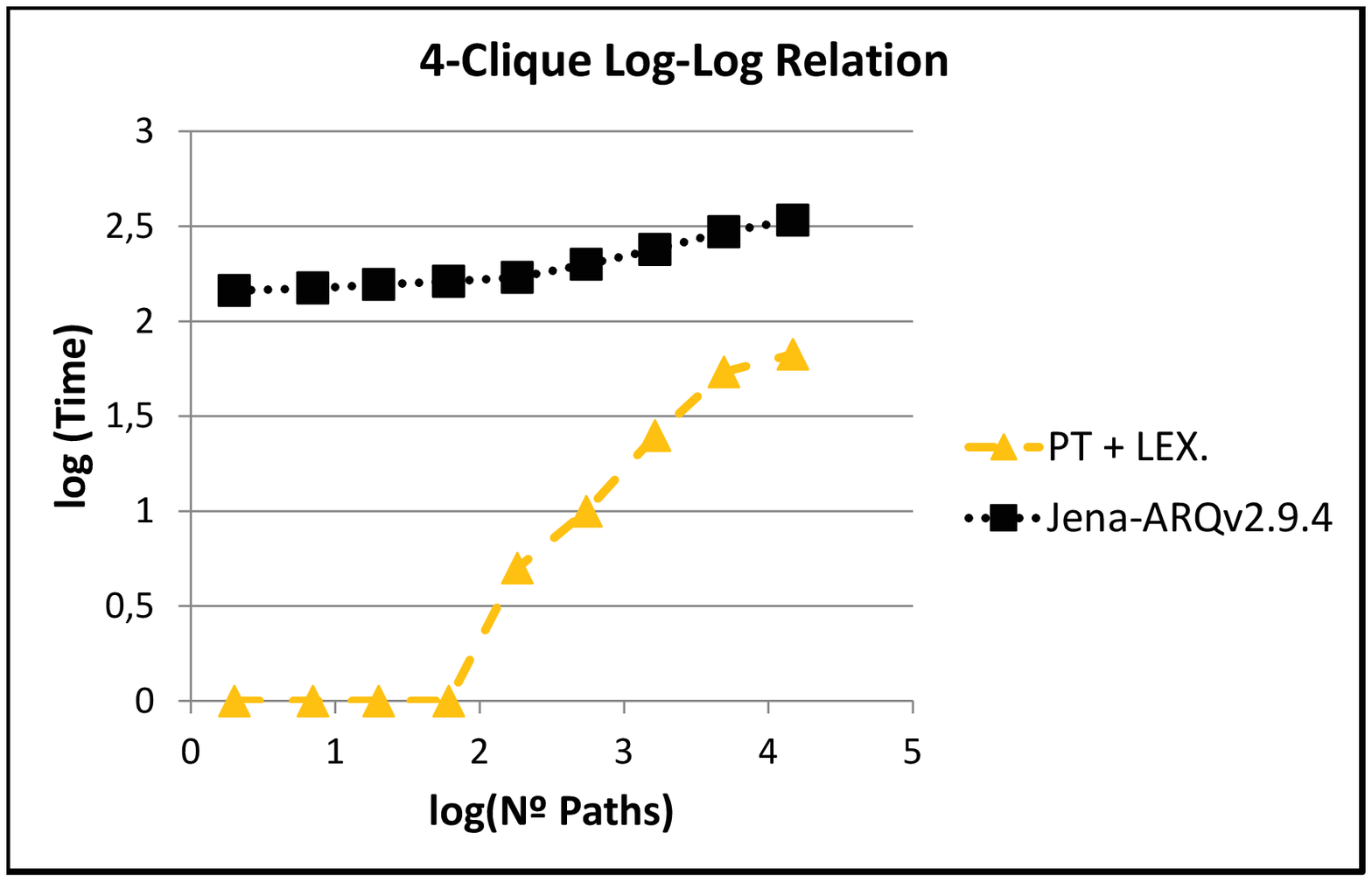}
        \caption{Time vs. number of paths log-log relation of 4-clique dataset.}
        \label{fig:loglog4clique}
    \end{subfigure}
     ~
    \begin{subfigure}[b]{0.45\textwidth}
        \includegraphics[width=\textwidth]{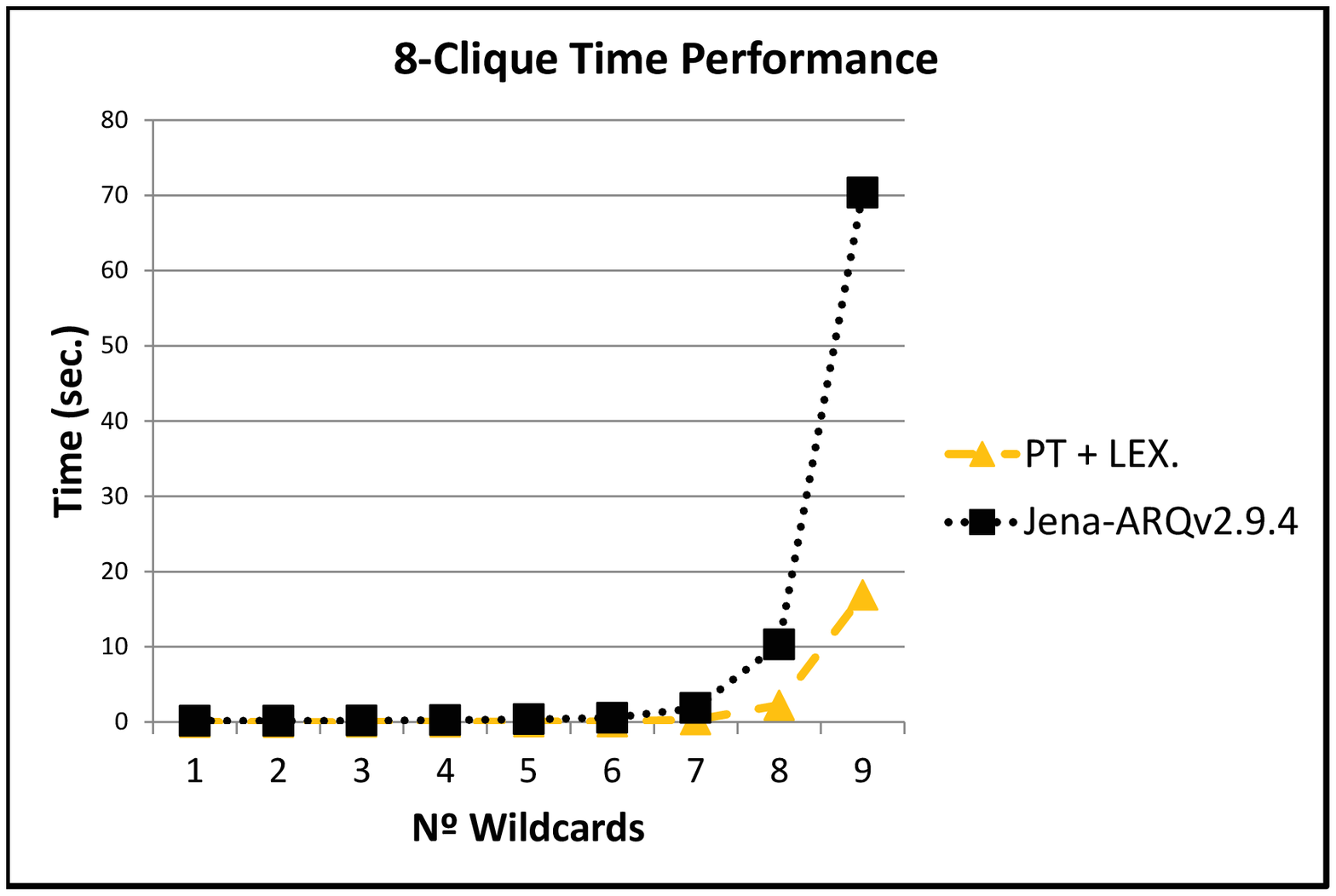}
        \caption{Time performance lookup operation of 8-clique dataset. }
        \label{fig:time8clique}
    \end{subfigure}
     ~
    \begin{subfigure}[b]{0.45\textwidth}
        \includegraphics[width=\textwidth]{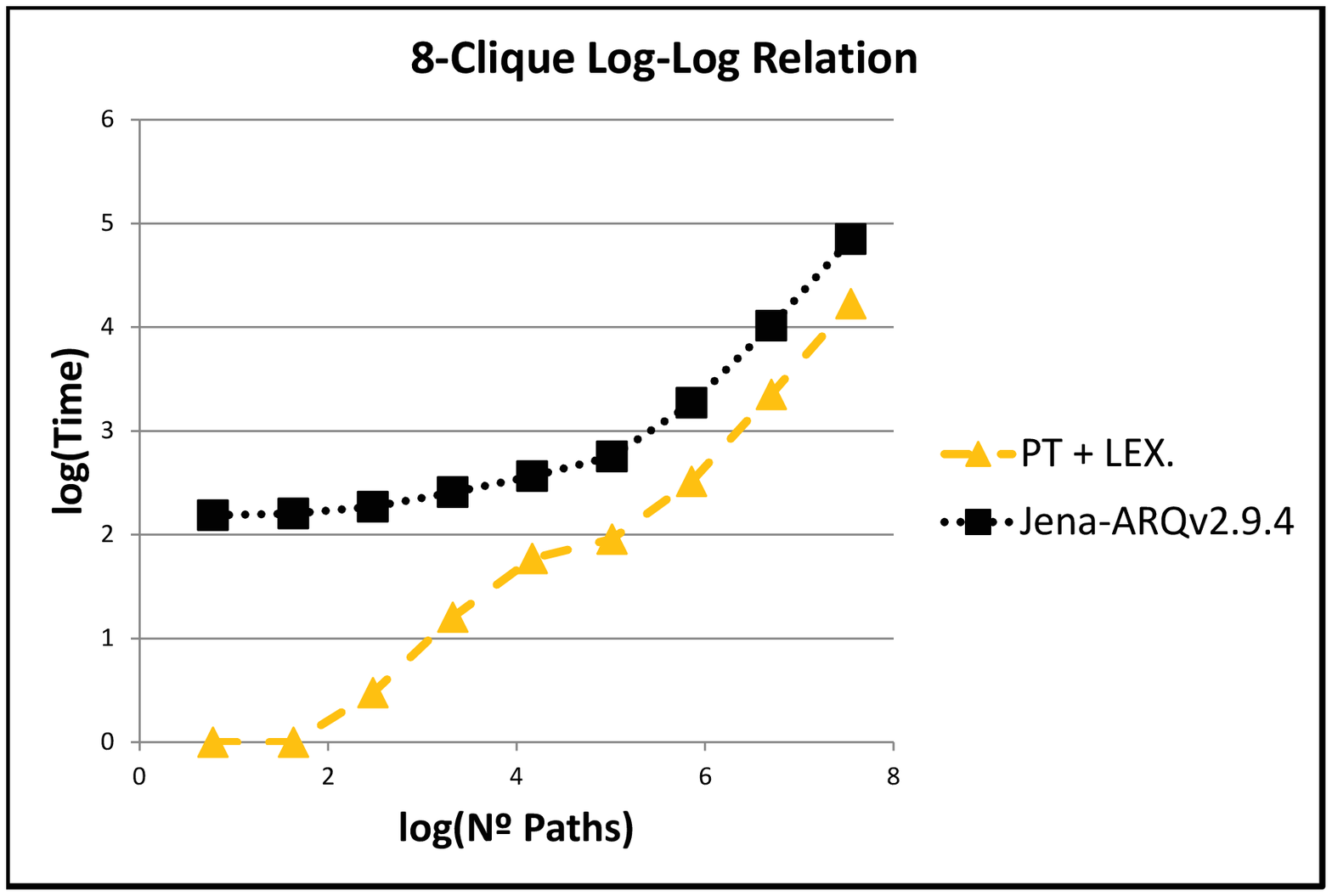}
        \caption{Time vs. number of paths log-log relation of 8-clique dataset.}
        \label{fig:loglog8clique}
    \end{subfigure}
\caption{E2 experiment results. On the left: time performance plot for ProvBench similar lookup application, 4-clique, and 8-clique. On the right: log-log plot showing the relation between time and the number of paths for ProvBench similar lookup application, 4-clique, and 8-clique configurations. All the plots contain two series, one for our approach PT+LEX. and another one for Jena-ARQv2.9.4 property paths.}\label{fig:experiments}
\end{figure}


Figure~\ref{fig:experiments} shows the results obtained for each dataset. The queries performed are of the type 'SELECT * WHERE \{ :r0 (:p)* :r1\}' for different number of wildcards path depth, ranging from 1 to 9. The main characteristic of the ProvBench scenario is the low number of linked processes, the main reason being the structure of the workflow is mainly linear and does not have cycles, as opposed to the 4/8-clique scenarios, which are fully connected graphs. On the other hand, ProvBench has a larger number of different processes than the other two scenarios.

\begin{figure}[H]
\centering
\hspace{-0.7cm}\epsfig{file=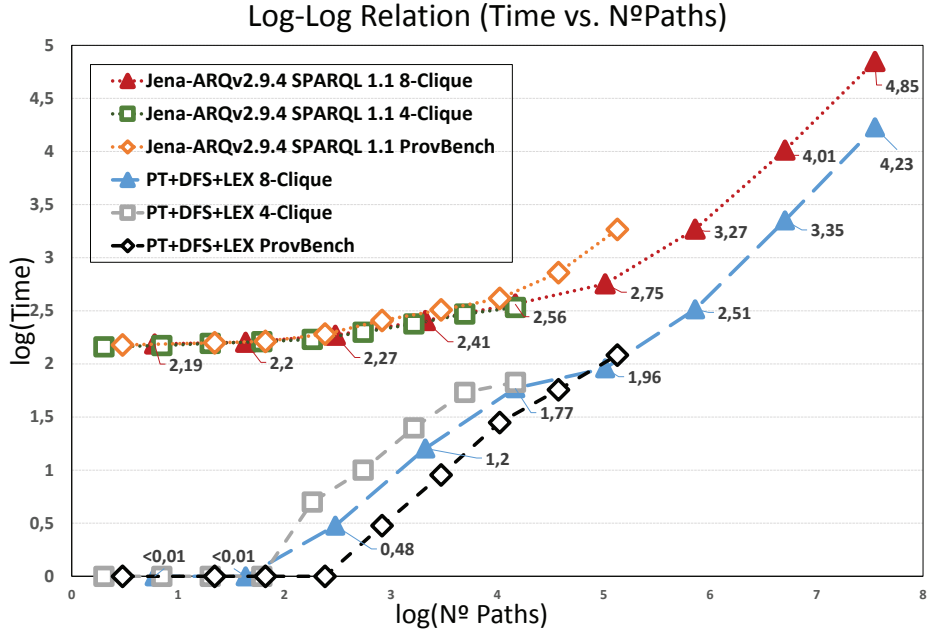, scale=0.45}
\caption{Overall log-log performance showing the performance on time vs. number of possible paths for each dataset and method(SPARQL1.1 and PT+DFS+LEX).}
\label{fig:globalperformance}
\end{figure}

Despite such differences, the behavior of the three scenarios is very similar as can be seen at the log-log plot on the right of the picture (being the x-axis the number of retrieved paths and the y-axis the time needed for executing the query) with the main difference on the slope of the curves, which is lower for the 4-clique scenario, followed by ProvBench, and by the 8-clique. It is worth highlighting that in the worst or more complex case (8-clique) the slope is lower than 1, meaning that the relation between the time and the complexity of the query is almost linear, which is the best expected situation. Therefore we can confirm that the implemented algorithm performs efficiently for both cases. For PT+DFS+LEX and Jena-ARQv2.9.4 the algorithms scale at worst linearly, being the number of retrieved solutions/paths the same for all the queries performed. However the left part of the Figure~\ref{fig:experiments} indicates that the time needed for performing the different queries is always larger for Jena-ARQv2.9.4 than for PT+LEX ($\approx$  270 ms., 1720 ms., and 54 secs. larger for 4-clique, ProvBench, and 8-clique, respectively), which shows an important improvement regarding time performance over using SPARQL 1.1. property paths. A summary of the log-log relation results for the E2 experimentation is shown in the Figure~\ref{fig:globalperformance}. 

As shown, for log(NºPaths) $\le$ 3.5 there is a large gap between the use of our approach or Jena-ARQ SPARQL1.1 that seems to have an offset since the beginning. Although we don't have the technical characteristics of the Jena-ARQ implementation it seems that there is an initial load acting as a common baseline for all the experiments. Furthermore we can verify within the range $6 < $ log(NºPaths) $< 7.5$ that this difference stabilizes at a factor $\approx$ of $\frac{1}{9}$ in favor of the presented approach. 

We also compared these results with the ones provided by the authors at~\cite{marcelo2012}, which presented results for using Psparql for 1, 2, and 3 complete cliques vs. using property paths obtaining the lowest execution performance time for the query 'SELECT DISTINCT * WHERE \{ :a0 (((:p)*)*)* :a1 \}'. However, their performance time (140 ms) is still worse than the results we obtained for the same query in a larger 8-clique graph.


\begin{table*}
\small 
\centering
\begin{tabular}{|p{3cm}|p{3cm}|p{3cm}|p{3cm}|}
\hline
\multicolumn{4}{|c|}{ProvBench Alternative Paths comparative}\\\hline
\hline
Number of wildcards & NºPaths ProvBench & NºPaths 8-Clique & \% ProvBench vs. 8-Clique									
\\\hline  
1 & 3  &  6    & 50.0\%  \\\hline
2	& 22 &  43   & 51.2\%  \\\hline
3	& 66 &  300  & 22.0\%  \\\hline
4 & 240 & 2101 & 11.4\%  \\\hline
5 & 823  &  14706    & 5.6\%  \\\hline
6	& 2953 &  102943   & 2.9\%  \\\hline
7	& 10483 &  720600  & 1.4\%   \\\hline
8 & 37480 & 5044201 & 0.7\%    \\\hline
9 & 133746 & 35309406 & 0.4\%  \\\hline
\end{tabular}
\caption{\small Comparative of the obtained number of alternative paths for ProvBench database and a 8-Clique graph.}
\label{table:provbench}
\end{table*}

For better understanding  of the Figure~\ref{fig:globalperformance} the table~\ref{table:provbench} shows a comparative between the number of alternative paths for different number of wildcards providing a short summary of the estimated link density of the ProvBench dataset vs. a complete 8-Clique graph.

\section{Conclusions}\label{sec:conclusions}
In this paper we presented a new method for transforming the graph structure associated to the provenance of workflow executions into a trie structure that enables indexing and fast accessibility. Such structure allows the characterization of workflows, not only based on their specification or the (often scarce) metadata associated to the resources related to a workflow, but on their execution behavior. 

Our method has been implemented using a variation of the generalized prefix tree and includes the canonical label definition as an important preprocessing step. The main application of the proposed indexing algorithm is to find in real time similar scientific workflows by identifying motifs, based on their execution traces, which share commonalities and could therefore be proposed as alternative means to perform a specific scientific computation related to an experiment or observation. The indexes and the statistics needed for obtaining the patterns and maximum likelihood values are calculated online providing tools for efficient querying. 
%
%


We conducted empirical evaluations of our approach both in terms of functionality and time performance, comparing our results against the possible number of alternative paths and the SPARQL 1.1. Property Paths W3C specification. We defined specific cases where our approach extends the functionality of the previous state of the art applications and we also evaluated our approach in more general scenarios such as complete graphs. Furthermore, we obtained improvements over the use of the SPARQL 1.1 Property Paths recommendation, hence providing a possible solution to its computational efficiency problem.
%
%

Finally, next steps include the application of our indexing structure in domain-specific repositories of scientific information, e.g. ROHub, rich in provenance data resulting from workflow executions in domains covering both experimental and observational disciplines. The resulting capabilities related to real-time querying of provenance information at scale are expected to further enhance the management of scientific data and methods as research objects, contributing to increase sharing and reuse. We will also seek to demonstrate the benefits of our approach in general-purpose semantic repositories in the Linked Open Data cloud as a suitable alternative to SPARQL 1.1 Property Paths. 

%
%

%
%
%
%
%

\section*{Acknowledgement} We gratefully acknowledge funding from the EU Horizon 2020 for research infrastructures and FP7 programs under grants EVER-EST-674907 and Wf4Ever-270129. Special thanks also to our University colleague Javier Martinez for his helpful discussions and insights. 

\bibliography{mybibfile}

\begin{thebibliography}{10}
\expandafter\ifx\csname url\endcsname\relax
  \def\url#1{\texttt{#1}}\fi
\expandafter\ifx\csname urlprefix\endcsname\relax\def\urlprefix{URL }\fi
\expandafter\ifx\csname href\endcsname\relax
  \def\href#1#2{#2} \def\path#1{#1}\fi

\bibitem{deelman2009workflows}
E.~Deelman, D.~Gannon, M.~Shields, I.~Taylor, Workflows and e-science: An
  overview of workflow system features and capabilities, Future generation
  computer systems 25~(5) (2009) 528--540.

\bibitem{doi:10.1093/nar/gkt328}
K.~Wolstencroft, R.~Haines, D.~Fellows, A.~Williams, D.~Withers, S.~Owen,
  S.~Soiland-Reyes, I.~Dunlop, A.~Nenadic, P.~Fisher, J.~Bhagat, K.~Belhajjame,
  F.~Bacall, A.~Hardisty, A.~Nieva de~la Hidalga, M.~P. Balcazar~Vargas,
  S.~Sufi, C.~Goble, The taverna workflow suite: designing and executing
  workflows of web services on the desktop, web or in the cloud, Nucleic Acids
  Research 41~(W1) (2013) W557--W561.

\bibitem{Ludascher:2006:SWM:1148437.1148454}
B.~Lud\"{a}scher, I.~Altintas, C.~Berkley, D.~Higgins, E.~Jaeger, M.~Jones,
  E.~A. Lee, J.~Tao, Y.~Zhao, Scientific workflow management and the kepler
  system: Research articles, Concurr. Comput. : Pract. Exper. 18~(10) (2006)
  1039--1065.

\bibitem{Gil:2007:WPC:1620113.1620127}
Y.~Gil, V.~Ratnakar, E.~Deelman, G.~Mehta, J.~Kim, Wings for pegasus: Creating
  large-scale scientific applications using semantic representations of
  computational workflows, in: Proceedings of the 19th National Conference on
  Innovative Applications of Artificial Intelligence - Volume 2, IAAI'07, AAAI
  Press, 2007, pp. 1767--1774.

\bibitem{myexperiment}
D.~De~Roure, C.~Goble, R.~Stevens, The design and realisation of the
  myexperiment virtual research environment for social sharing of workflows.,
  Future Generation Computer Systems 25 (2009) 561--567.

\bibitem{crowdlabs11}
P.~Mates, E.~Santos, J.~Freire, C.~T. Silva, Crowdlabs: Social analysis and
  visualization for the sciences, In 23rd International Conference on
  Scientific and Statistical Database Management (SSDBM) (2011) 555--564.

\bibitem{gal11}
J.~Goecks, A.~Nekrutenko, J.~Taylor, Galaxy: a comprehensive approach for
  supporting accesible, reproducible, and transparent computational research in
  the life sciences., Genome Biol 11(8): R86 (2011)  .

\bibitem{cohen11}
S.~Cohen-Boulakia, U.~Leser, "search, adapt, and reuse: The future of
  scientific workflows, In International Conference on Management Data
  (SIGMOD'11) 20 (2011) 6--16.

\bibitem{Getoor03}
L.~Getoor, Link mining: a new data mining challenge, SIGKDD Explor. Newsl.
  5~(1) (2003) 84--89.

\bibitem{zhang12}
X.~Zhang, C.~Zhao, P.~Wang, F.~Zhou, Mining link patterns in linked data,
  Web-Age Information Management, LNCS 7418 (2012) 83--94.

\bibitem{knuth97}
D.~E. Knuth, The art of computer programming Volume I. Fundamental Algorithms
  (Third Edition), Addison-Wesley, New York, 1997.

\bibitem{fredkin60}
E.~Fredkin, Trie memory, Communications of the ACM 3~(9) (1960) 490--499.

\bibitem{de702af8-ea61-47af-aabb-286b2274e1be}
P.~Amstutz, N.~Tijanić, S.~Soiland-Reyes, J.~Kern, L.~Stojanovic, T.~Pierce,
  J.~Chilton, M.~Mikheev, S.~Lampa, H.~Ménager, S.~Frazer, V.~S. Malladi,
  M.~R. Crusoe, Portable workflow and tool descriptions with the cwl (common
  workflow language), F1000Research.

\bibitem{Bechhofer2013599}
S.~Bechhofer, I.~Buchan, D.~D. Roure, P.~Missier, J.~Ainsworth, J.~Bhagat,
  P.~Couch, D.~Cruickshank, M.~Delderfield, I.~Dunlop, M.~Gamble,
  D.~Michaelides, S.~Owen, D.~Newman, S.~Sufi, C.~Goble, Why linked data is not
  enough for scientists, Future Generation Computer Systems 29~(2) (2013) 599
  -- 611.

\bibitem{8109145}
J.~M. Gomez-Perez, R.~Palma, A.~Garcia-Silva, Towards a human-machine
  scientific partnership based on semantically rich research objects, in: 2017
  IEEE 13th International Conference on e-Science (e-Science), 2017, pp.
  266--275.

\bibitem{palma2014rohub}
R.~Palma, P.~Ho{\l}ubowicz, O.~Corcho, J.~Gomez-Perez, C.~Mazurek, Rohub—a
  digital library of research objects supporting scientists towards
  reproducible science, Semantic Web Evaluation Challenge (2014) 77--82.

\bibitem{Zhao2012WhyWB}
J.~Zhao, J.~M. G{\'o}mez-P{\'e}rez, K.~Belhajjame, G.~Klyne,
  E.~Garc{\'i}a-Cuesta, A.~Garrido, K.~M. Hettne, M.~Roos, D.~D. Roure, C.~A.
  Goble, Why workflows break — understanding and combating decay in taverna
  workflows, 2012 IEEE 8th International Conference on E-Science (2012) 1--9.

\bibitem{garijo2014common}
D.~Garijo, P.~Alper, K.~Belhajjame, O.~Corcho, Y.~Gil, C.~Goble, Common motifs
  in scientific workflows: An empirical analysis, Future Generation Computer
  Systems 36 (2014) 338--351.

\bibitem{moreau10}
L.~Moreau, The foundations for provenance on the web, Found. Trends Web Sci.
  2~(2-3) (2010) 99--241.

\bibitem{1316839}
W.~van~der Aalst, T.~Weijters, L.~Maruster, Workflow mining: discovering
  process models from event logs, IEEE Transactions on Knowledge and Data
  Engineering 16~(9) (2004) 1128--1142.

\bibitem{mor08}
L.~Moreau, B.~Ludaescher, e.~a. Altintas, Special issue: The first provenance
  challenge., Concurrency and Computation: Practice and Experience 20 (2008)
  409--418.

\bibitem{cohen09}
S.~Cohen-Boulakia, W.-C. Tan, Provenance in scientific databases, Encyclopedia
  of Database Systems (2009) 2202--2207.

\bibitem{Alper:2014:LEW:2977935.2977945}
P.~Alper, K.~Belhajjame, C.~A. Goble, P.~Karagoz, Labelflow: Exploiting
  workflow provenance to surface scientific data provenance, in: Revised
  Selected Papers of the 5th International Provenance and Annotation Workshop
  on Provenance and Annotation of Data and Processes - Volume 8628, IPAW 2014,
  Springer-Verlag New York, Inc., New York, NY, USA, 2015, pp. 84--96.

\bibitem{6655668}
A.~Chebotko, J.~Abraham, P.~Brazier, A.~Piazza, A.~Kashlev, S.~Lu, Storing,
  indexing and querying large provenance data sets as rdf graphs in apache
  hbase, in: 2013 IEEE Ninth World Congress on Services, 2013, pp. 1--8.

\bibitem{Getoor05}
L.~Getoor, C.~P. Diehl, Link mining: a survey, SIGKDD Explor. Newsl. 7~(2)
  (2005) 3--12.

\bibitem{boehm11}
M.~Boehm, B.~Schlegel, P.~Benjamin, U.~Fischer, Efficient in-memory indexing
  with generalized prefix trees, BTW 180 (2011) 227--246.

\bibitem{brenes08}
S.~Brenes, Y.~Wu, D.~Gucht, P.~S. Cruz, Trie indexes for efficient xml query
  evaluation, Proceedings of the 11th International Workshop on Web and
  Databases WebDB (2008)  .

\bibitem{morariu11}
D.~I. Morariu, R.~G. Cre{\'N}ulescu, L.~N. Vin{\'N}an, Using suffix tree
  document representation in hierarchical agglomerative clustering,
  International Conference on Intelligent Systems--ICIS Conference, Paris
  (2011)  .

\bibitem{DBLP:journals/eswa/Garcia-CuestaI12}
E.~Garc\'{\i}a-Cuesta, J.~A. Iglesias, User modeling: Through statistical
  analysis and subspace learning, Expert Syst. Appl. 39~(5) (2012) 5243--5250.

\bibitem{Grig08}
D.~Grigori, J.~C. Corrales, M.~Bouzeghoub, Behavioral matchmaking for service
  retrieval: Application to conversation protocols, Inf. Syst. 33~(7-8) (2008)
  681--698.

\bibitem{he08}
H.~He, A.~K. Singh, Graphs-at-a-time: query language and access methods for
  graph databases, In the proceedings of the SIGMOD conference, Vancouver,
  Canada (2008) 405--418.

\bibitem{heinz02}
S.~Heinz, J.~Zobel, H.~E. Williams, Burst tries: A fast, efficient data
  structure for string keys, ACM Transactions on Information Systems 20 (2002)
  192--223.

\bibitem{tobin86}
T.~J. Lehman, M.~J. Carey, A study of index structures for main memory database
  management systems, Proceedings of the 12th International Conference on Very
  Large Data Bases (1986) 294--303.

\bibitem{marcelo2012}
M.~Arenas, S.~Conca, J.~P{\'e}rez, Counting beyond a yottabyte, or how sparql
  1.1 property paths will prevent adoption of the standard, Proceedings of the
  21st international conference on World Wide Web (2012) 629--638.

\bibitem{xifeng02}
Y.~Xifeng, H.~Jiawei, gspan: Graph-based substructure pattern mining, In the
  proceedings of the IEEE International Conference of Data Mining (ICDM) (2002)
  721--724.

\bibitem{hayes04}
P.~Hayes, {RDF} semantics. {W3C} recommendation (February 10, 2004).

\bibitem{kha13}
K.~Belhajjame, et~al., A workflow prov-corpues based on taverna and wings,
  Proceedings of the Joint EDBT/ICDT 2013 Workshops (2013) 331--332.

\bibitem{miss10}
P.~M. et~al., Wings: Intelligent workflow-based design of computational
  experiments. taverna, reloaded., Proceedings of SSDBM'10 (2010)  .

\bibitem{gil11}
Y.~Gil, V.~Ratnakar, J.~Kim, et~al., Wings: Intelligent workflow-based design
  of computational experiments, IEEE Intelligent Systems 1~(26) (2011) 62--72.

\end{thebibliography}

\end{document}